\documentclass[aps,pra,reprint]{revtex4-1}

\usepackage{bm}

\begin{document}

\title{Atomic Two-Body $\bm{(Z\alpha )^6}$ Predictions } 

\author{John H. Connell}

\email{connell@stcc.edu}

\affiliation{Springfield Technical Community College\\
Springfield, Massachusetts, USA}

\date{\today}

\begin{abstract}

For muonic hydrogen, positronium, and ordinary hydrogen, we show that the 
existence of a relativistic two-body wave equation whose energy levels are 
physically accurate to order $(Z\alpha )^4$ (where $Z\alpha$ is the binding coupling constant) 
implies a previously unknown two-body Sommerfeld energy formula which can be 
used to predict energy terms to order $(Z\alpha )^6$ using simple algebra.  Two such 
terms are verified to be physically correct by earlier $(Z\alpha 
)^6$ calculations for 
positronium.  For muonic hydrogen and ordinary hydrogen, these terms are 
predictions.  

\end{abstract}

\pacs{31}

\maketitle


\section{Introduction}

Although the Dirac equation for a single fermion in a static Coulomb field was 
solved analytically in 1928 \cite{1928}, including the derivation of the 
single-particle Sommerfeld formula expressing the energy levels to any order in 
the binding coupling constant $Z\alpha$, a comparable solution has never been found for 
atomic two-particle systems such as muonic hydrogen, positronium, and ordinary 
hydrogen.  For these atomic systems, energy levels have always been calculated 
using first-order perturbation theory to obtain energy levels to order $
(Z\alpha )^4$, and 
second-order perturbation theory to find energy levels to order $
(Z\alpha )^6$.

In this note we demonstrate a possible first step towards an analytic solution to 
the atomic two-body bound-state problem.  It is shown that a two-body 
relativistic wave equation which predicts physically accurate energy levels to 
order $(Z\alpha )^4$ also leads to a new, two-body, Sommerfeld energy-level formula 
which, by simple algebra, predicts two energy-level terms in order $
(Z\alpha )^6$ which 
have already been found to be physically correct in positronium by various 
authors.  These terms constitute predictions for ordinary hydrogen and muonic 
hydrogen.

We start with the new predictions.  Afterwards we will state the two-body 
relativistic wave equation and show how it leads to the two-body Sommerfeld 
formula.

\section{Predictions}

The simple condition that the two-body relativistic wavefunction exists leads 
directly to what we believe is the first known two-body Sommerfeld energy 
formula:  
\begin{equation}E=\sqrt {m^2+M^2+\frac {2mM}{\sqrt {1+\frac {
(Z\alpha )^2}{(N+\Delta\epsilon )^2}}}}\label{E}\end{equation}
Here $m$ and $M$ are the masses of the bound particles and  $
N$  is the Bohr 
quantum number. The  quantity
\begin{equation}\Delta\epsilon\:=\:(Z\alpha )^2\epsilon_2+(Z
\alpha )^4\epsilon_4+(Z\alpha )^6\epsilon_6+\cdots\label{Deps}\end{equation}
is a function of the total system spin $F$, and of mixtures of $
L$ and $S$.  The 
leading coefficient $\epsilon_2$ is evaluated below.

With the usual definition  $\mu =$$mM/(m+M)$, and using the abbreviation 
$x=\mu /(m+M)$, the expansion of (\ref{E}) in powers of $(Z\alpha 
)^2$  gives
\begin{equation}E=m+M+C_2\:(Z\alpha )^2\mu +C_4\:(Z\alpha )^4\mu +C_6\:(Z\alpha )^6\mu  +\cdots\label{
Eexpn}\end{equation}
in which
\begin{eqnarray}
C_2&=&-\,\frac 1{2N^2}\label{C2}\\
C_4&=&\frac {3-x}{8N^4}+\frac {\epsilon_2}{N^3}\label{C4}\\
C_6&=&-\,\frac {5-3x+x^2}{16N^6}-\frac {(3-x)\epsilon_2}{2N^
5}-\frac {3\epsilon_2^2}{2N^4}+\frac {\epsilon_4}{N^3}\label{C6}
\end{eqnarray}
Equation  (\ref{C4}) shows that the leading coefficient $\epsilon_
2$ in (\ref{Deps}) can be 
read off from an ordinary first-order-perturbation calculation of the energy.  This 
will be done in Sec.  III, in which $\epsilon_2$ is shown as eqn.  (\ref{eps2}).  Therefore $
\epsilon_2$ 
is known.  

Equation (\ref{C6}) contains four $(Z\alpha )^6$ energy terms, three of which can be 
predicted since $\epsilon_2$ is known.  We shall see that two of these simple predictions 
agree with far more complicated perturbation calculations in the case of 
positronium, whose $(Z\alpha )^6$ terms have already been worked out by various authors.  
We will discuss why.

We now review the predictions and report on  their verification 
for positronium.

\subsection{Prediction for  $\bm{(Z\alpha )^6/N^6}$   }

Equation (\ref{C6}) predicts that the $(Z\alpha )^6/N^6$ energy term due to the binding 
interaction $Z\alpha$ of   atomic two-body bound states is 
\begin{equation}-\:\frac {5-3x+x^2}{16}\:\frac {(Z\alpha )^6}{
N^6}\:\mu\label{N6}\end{equation}
for every $N$, and every angular  quantum number, including those of $
L=0$.

\subsection{Verification for Positronium}

For positronium, in which $m=M$,  $\mu =m/2$, and $x=1/4$, it is easy to see that 
for every angular state of positronium, equation (\ref{N6}) predicts that
the $(Z\alpha )^6/N^6$ term of the energy will always be
\begin{equation}-\frac {69}{512}\:\frac {(Z\alpha )^6}{N^6}\:
m\label{69512}\end{equation}

Precisely this value  was found for all P-states by Khriplovich et al.  
\cite{P}, for both S-states by Czarnecki et al.  \cite{Cz}, and then in all $
L\ge 2$ states 
in Zatorski's recent calculations \cite{Jacek}.  

To contrast the methods, the predictions (\ref{N6}) and (\ref{69512}) come from the 
simple algebraic expansion of the two-body Sommerfeld formula (\ref{E}).  The 
calculations of (\ref{69512})   in standard perturbation theory   by Zatorski are the 
sum of five terms.  Three terms are from second-order perturbation theory (eqns.  
(159), 160) and (162) of ref.  \cite{Jacek}).  Two terms are first-order (eqns.  (91) 
and (122) of ref.  \cite{Jacek}). One contains the expectation value of $
{\bf p}^{{\bf 6}}$.  

Equation (\ref{N6}) is a prediction of the results of future calculations on 
ordinary hydrogen and muonic hydrogen.

\subsection{Prediction for  $\bm{(Z\alpha )^6/N^4}$   }

Equation (\ref{C6}) with the coefficient $\epsilon_2$ given in eqn.  (\ref{eps2}) below 
predicts that the $(Z\alpha )^6/N^4$ energy term of  atomic two-body bound states is 
\begin{equation}-\frac 32\:\epsilon^2_2\:\:\frac {(Z\alpha )^
6}{N^4}\:\mu\label{N4}\end{equation}

\subsection{Verification for Positronium}

As an example, for $L=F+1$, the coefficient $\epsilon_2$ in eqn. (\ref{eps2}) below is
\[\epsilon_2\:(\mbox{\rm Positronium, }L=F+1)\:=\:-\frac 12\:\left
[\frac 1L+\frac 1{(2L+1)(2L-1)}\right]\]
In eqn.  (211) of Zatorski  \cite{Jacek} the calculated coefficient of the $
(Z\alpha )^6m/N^4$ 
term for $L=F+1$ is 
\[-\,\frac {3-6L-21L^2+24L^3+48L^4}{16L^2(2L-1)^2(2L+1)^2}\]
Bearing in mind that $\mu =m/2$, this is {\em exactly\/} the predicted term (\ref{N4}) above.  
This verifies the prediction of the two-body Sommerfeld formula for the 
positronium $L=F+1$ state.  Our prediction is also found to hold for the other 
three positronium states (eqns.  (207), (215) and (219) of ref.  \cite{Jacek}).   (For 
the $^1S_0$ and $^3S_1$ states of positronium, the coefficients of the $
(Z\alpha )^6m/N^4$ term 
contain other contributions (see ref.  \cite{Cz}) and no prediction can be made.)

To contrast the methods, our prediction comes from simple algebra using the 
$(Z\alpha )^4$ first-order perturbation result for $\epsilon_
2$.  In ref.  \cite{Jacek}, six 
second-order perturbation results had to be added together (eqns.  (153), (158), 
(164), (171), (174) and (177)).

Equation (\ref{N4}) is a prediction of the results of future calculations on 
ordinary hydrogen and muonic hydrogen.

\subsection{Prediction for  $\bm{(Z\alpha )^6/N^5}$   }

Equation (\ref{C6}) predicts that the $(Z\alpha )^6/N^5$ energy contribution due to the 
two-body wave equation is
\[-\:\frac {(3-x)\epsilon_2}2\:\frac {(Z\alpha )^6}{N^5}\:\mu\]
The $1/N^5$ terms of Zatorski \cite{Jacek} do {\em not\/} agree with this prediction.  This 
leads to a review of what an analytic solution of the relativistic two-body 
problem would provide.

We recall that different  theoretical starting points (for example the 
one-photon-exchange Bethe-Salpeter equation, and the conventional Breit equation) 
can lead to the same $(Z\alpha )^4$ energy levels but different $
(Z\alpha )^6$ levels.  In 
Hamiltonian language, with $H_0$ the Coulomb Schr\"odinger Hamiltonian, one can have 
$H_0+V_1$ and $H_0+V_2$ such that $\left<V_1\right>=\left<V_
2\right>$ to order $(Z\alpha )^4$, while $\left<V_1-V_2\right
>\neq 0$ to 
order $(Z\alpha )^6$.  However, it is easy to show that in this case 
$\left<V_1\frac 1{(E_0-H_0)'}V_1\right>=\left<V_2\frac 1{(E_
0-H_0)'}V_2\right>$ to order $(Z\alpha )^6$.  This means that the differences 
in the theories can be corrected using only {\em first\/}-order perturbation theory to 
order $(Z\alpha )^6$.  Thus if {\em any\/} theory, physically correct to order $
(Z\alpha )^4$, is able to 
calculate its energy levels to order $(Z\alpha )^6$ easily, the physically 
correct levels to order $(Z\alpha )^6$ can be calculated as first-order corrections only.  
That is one of the motivations for the present work.  

Examination of Zatorski's $(Z\alpha )^6$ results (ref.  \cite{Jacek}, Appendix A) shows that 
$1/N^6$ and $1/N^4$ appear in the second-order perturbation terms, while $
1/N^5$ and 
$1/N^3$ dominate in the first-order expectation values.  It is not surprising 
therefore that our $(Z\alpha )^6$ predictions for the $1/N^6$ and $
1/N^4$ terms are physically 
correct, while the $1/N^5$ and $1/N^3$ predictions would need first-order corrections.

\section{Two-Body Wave Equation}

The relativistic atomic two-body wave equation from which these results 
are obtained was derived in ref.  \cite{1991} from the Bethe-Salpeter equation for 
two spin-$1/2$  point particles bound by a single-photon-exchange kernel 
in the Coulomb gauge.  The derivation used a simple quasi-potential approximation 
\cite{1967,1971} with its associated Blankenbecler-Sugar correction series 
\cite{1966}.

The bound-state energy $E$ is parametrised by a quantity $\beta$ as follows:  
\begin{equation}E=\sqrt {m^2-\beta^2}+\sqrt {M^2-\beta^2}\label{Ebsq}\end{equation}
The particles' individual bound-state energies also occur:
\begin{equation}t=\sqrt {m^2-\beta^2}\mbox{\rm ,}\qquad T=\sqrt {
M^2-\beta^2}\label{tT}\end{equation}
The Pauli matrices $\bm{\sigma}$ and the Dirac matrices $\bm{
\gamma}$, $\gamma^0$ refer to the 
particle of mass $m$, while $\bm{\Sigma}$, $\bm{\Gamma}$ and $
\Gamma^0$ refer to the particle of mass 
$M$.  The operator ${\bf p}$ is $-i\bm{\nabla}$, where $\bm{
\nabla}$ refers to the relative 
position co\"ordinate ${\bf r}$.  Also $r=|{\bf r}|$ and $\hat {
{\bf r}}\equiv {\bf r}/r$.  The only constant of the motion 
is $\bm{F}=\bm{L}+\bm{\sigma}/2+\bm{\Sigma}/2$.

The relativistic bound-state wave equation in the centre-of-mass system is
\begin{eqnarray}
&&[{\bf p}^2+\beta^2]\psi ({\bf r})\nonumber\\*
&&=-\frac 1{2E}\left[m-\bm{\gamma}\cdot {\bf p}+\gamma^0t\right
]\left[M+\bm{\Gamma}\cdot {\bf p}+\Gamma^0T\right]\times\nonumber\\*
&&\times\left\{-\gamma^0\Gamma^0\:\frac {Z\alpha}r+\frac {\bm{
\gamma}\cdot\bm{\Gamma}+\bm{\gamma}\cdot\hat {{\bf r}}\:\bm{
\Gamma}\cdot\hat {{\bf r}}}2\:\frac {Z\alpha}r\right.+\label{we}\\*
&&+\left.\frac 1{2E}\left(\frac {Z\alpha}r\right)^2-\frac {g
-2}{4M}\bm{\gamma}\cdot\hat {\bf r}\times\bm{\Sigma}\:\frac {
Z\alpha}{r^2}\right\}\psi ({\bf r})\nonumber\end{eqnarray}
The eigenvalue is $\beta^2$, not $E$.  The constant $\beta^2$ is substituted into the square roots 
in equation (\ref{Ebsq}) to obtain the energy $E$.  There are no non-local operators of 
the form $\sqrt {{\bf p}^2+m^2}$.  No terms ${\bf p}^4$, ${\bf p}^
6$ appear in perturbation theory.  

The first and second terms in the curly brackets are the standard binding 
potential and Breit interaction.  The third term is $1/2E$ times the square of the 
binding potential.  This term is a consequence of the Blankenbecler-Sugar 
correction formalism. It ensures that the relativistic energies are correct to first 
order  \cite{SLAC}.

The final term in the curly brackets, which was not included in ref.  \cite{1991}, 
contains an anomalous magnetic moment for the particle of mass $
M$.  It is derived 
by adding a term $i(g-2)({\bf k}\times\bm{\Sigma})/4M$ to the Dirac matrix $\bm{
\Gamma}$ in the vertex 
function of the particle of mass $M$ in the originating Bethe-Salpeter equation (see 
e.g.  Carlson \cite{Carlson}, eqn.  (5)).  The term is included so that the particle 
of mass $M$ may represent a point proton.  

The wave equation (\ref{we}) has two singularities:  one at $
r=0$ as usual, and 
another at roughly $r=Z\alpha /2E$ due to the double derivative $\bm{
\gamma}\cdot {\bf p}\:\bm{\Gamma}\cdot {\bf p}$.  
When the positive particle is a proton the distance $Z\alpha 
/2E$ is about $0.001$ f, well 
inside the proton.  For positronium $Z\alpha /2E$ is about $
\alpha^2$ times the Bohr radius.  
The discussion below will be for larger radii than these.

To verify the correctness of the wave equation (\ref{we}) we give its bound-state 
energies to order $(Z\alpha )^4$.  Recalling that $x=\mu /(m
+M)$, from first-order 
perturbation theory the bound-state energies of the wave equation (\ref{we})  
to order $(Z\alpha )^4$ are
\begin{equation}E=m+M-\frac {(Z\alpha )^2}{2N^2}\:\mu +\frac {
3-x\:}8\:\frac {(Z\alpha )^4}{N^4}\:\mu +\epsilon_2\:\frac {
(Z\alpha )^4}{N^3}\:\mu\label{E4}\end{equation}
in which
\begin{equation}\epsilon_2=\left\{\begin{array}{ll}
-\frac 12\left[\frac 1L+\frac {2xg}{(2L+1)(2{\rm F}+1)}\right
]\quad&L={\rm F}+1\\
-\frac 12\left[\frac 1{L+1}-\frac {2xg}{(2L+1)(2{\rm F}+1)}\right
]\quad&L={\rm F}-1\\
-\frac 14\left[\frac 1L+\frac 1{L+1}+\frac {\sqrt {1+4a^2}}{
(2L+1)L(L+1)}\right]\quad&L={\rm F},S\approx 1\\
-\frac 14\left[\frac 1L+\frac 1{L+1}-\frac {\sqrt {1+4a^2}}{
(2L+1)L(L+1)}\right]\quad&L={\rm F},S\approx 0\end{array}
\right.\label{eps2}\end{equation}
where the quantity $a^2$ is 
\begin{equation}a^2=\left[\frac {\mu}M-\frac {\mu}m+(g-2)x\right
]^2L(L+1)\label{asq}\end{equation}

The symbols $S\approx 1$, $S\approx 0$ for $L={\rm F}$ stand for the state in which $
S$ is 
predominantly $1$ or $0$, respectively.  This expression for $
\epsilon_2$ is correct for $L=0$ (in 
the case of positronium, it does not include the annihilation term), but for clarity 
we give its values for $L=0$ explicitly:  
\begin{equation}\epsilon_2\:(L=0)=\left\{\begin{array}{ll}
-\frac 12+xg/3&\quad S=1\\
-\frac 12-xg&\quad S=0\end{array}
\right.\label{eps2L0}\end{equation}

Equations (\ref{Deps}), (\ref{C4}), and (\ref{E4}) confirm that $
\epsilon_2$ is the 
first coefficient in the expansion of the small angular parameter $
\Delta\epsilon$.

Referring back to standard references, we find that the energy levels (\ref{E4}) to 
order $(Z\alpha )^4$ agree with all known cases.  

For example, it is easy to see that the Dirac-Coulomb limit $
M\rightarrow\infty$ (where $x=0$) 
has the correct fine structure:  $\epsilon_2$ is always $-1/
2(j+\frac 12)$.  For positronium, with 
$m=M$ and $g=2$, we find the standard energy levels, without the annihilation 
term (Bethe and Salpeter \cite{BS}, Sec.  23).

The hyperfine splittings are also correct to order $m/M$ (Bethe \& Salpeter 
\cite{BS}, Sec.  22, or White \cite{White}, Sec.  18.3).  (Note that in the literature 
the factor $Z$ is not included in the magnetic moment of the proton, so $
Z$ appears 
cubed.  Here for consistency we carry $Z$, which is one, to the fourth power.)  For 
muonic hydrogen, using $\Delta\epsilon =(Z\alpha )^2\epsilon_
2$ in the Sommerfeld formula (\ref{E}), we find 
that the energy difference between the standard hyperfine levels (first-order in 
$m/M$), and the Sommerfeld formula, is at most $0.005$ meV.

It only remains to obtain the two-body Sommerfeld energy-level formula (\ref{E}).

\section{Derivation of the  Two-Body  Sommerfeld Formula}

Following conventional treatments of the Coulomb Schr\"odinger equation and the 
Coulomb Dirac equation, we substitute 
\begin{equation}\psi ({\bf r})=e^{-\beta r}r^{\epsilon}\sum_
ja_jr^j\label{wfn}\end{equation}
into the wave equation (\ref{we}).  Here the coefficients $a_
j$ are 16-dimensional 
vectors.  The expansion is expected to be valid for $r\gg Z\alpha 
/2E$.  One obtains a 
four-term recurrence relation for the coefficients $a_j$.  

When the dominant terms acting on the large component of the wave function in 
the wave equation (\ref{we}) are examined, we see that they are the same as in 
the Coulomb Schr\"odinger equation.  That means that if the series (\ref{wfn}) does 
not terminate for some $j=n$, the wavefunction will diverge as $
e^{+\beta r}$ for large $r$.  
So that the wavefunction (\ref{wfn}) will exist,  we 
 assume that the series terminates at $j=n$.  Then it is easy to find that 
with $a_j=0$ for $j>n$, the recurrence relation gives this equation for $
a_n$:  
\begin{eqnarray}
&&2\beta (\epsilon +1+n)a_n\nonumber\\
&=&-\frac {Z\alpha}{2E}\tilde {m}\tilde {M}\left[-\gamma^0\Gamma^
0+\frac {\bm{\gamma}\cdot\bm{\Gamma}+\bm{\gamma}\cdot\hat {{\bf r}}\:\bm{
\Gamma}\cdot\hat {{\bf r}}}2\right]a_n\label{an}\end{eqnarray}
containing the projection operators
\begin{equation}\tilde {m}=m-i\beta\bm{\gamma}\cdot\hat {{\bf r}}
+\gamma^0t,\qquad\tilde {M}=M+i\beta\bm{\Gamma}\cdot\hat {{\bf r}}
+\Gamma^0T.\label{mM}\end{equation}

To solve equation (\ref{an}), we note that $a_n$ has the form
 $a_n=\tilde {m}\tilde {M}b.$ Substituting that back into (\ref{an}) 
puts   $\tilde {m}\tilde {M}$ on each side of the Coulomb and Breit terms. 
Multiplying them out gives  a scalar multiple of $\tilde {m}
\tilde {M}$ again.  One thus finds 
\[2\beta (\epsilon +n+1)a_n=-\frac {Z\alpha}{2E}\left[-4tT+4
\beta^2\right]a_n\nonumber\]
which is to say
\begin{equation}\frac {\beta E}{tT-\beta^2}=\frac {Z\alpha}{
\epsilon +n+1}\label{bE}\end{equation}

In the case of the one-particle Dirac-Coulomb equation  it is well known that
\begin{equation}n+\epsilon+1=N\:-\:\frac {(
Z\alpha )^2}{(j+\frac 12)+\sqrt {(j+\frac 12)^2-(Z\alpha )^2}}\label{DCoul}\end{equation}
This example suggests that in the two-body equation (\ref{bE}) 
\begin{equation}n+\epsilon +1=N+\Delta\epsilon\label{NDe}\end{equation}
with $\Delta\epsilon$ the expansion in powers of $(Z\alpha )^
2$  shown in equation (\ref{Deps}).  With 
this assumption, using equations (\ref{Ebsq}) and (\ref{tT}), equation (\ref{bE})  
immediately gives the two-body Sommerfeld energy formula (\ref{E}).

In conclusion, the surprising discovery of a Sommerfeld energy-level formula for 
two-body atoms, which predicts $(Z\alpha )^6$ energy terms two of which are verified to 
be physically correct for positronium, allows the hope that one day it may be 
possible to find an analytic solution to the atomic two-body bound-state problem 
analogous to the one-particle solution of 1928.

\end{document}